\documentclass[journal=jacsat,manuscript=article]{achemso}

\usepackage[version=3]{mhchem} 

\usepackage{siunitx}
\usepackage{graphicx}
\usepackage{subcaption}
\usepackage{float}

\graphicspath{ {./images/} }

\DeclareSIUnit\angstrom{\text {Å}}

\usepackage{soul}
\DeclareSIUnit\cal{cal}
\DeclareSIUnit\kcal{\kilo\cal}

\author{Jeet Majumdar}
\affiliation{Centre for Condensed Matter Theory, Department of Physics, Indian Institute of Science, Bangalore 560012, India}

\author{Shubhadeep Nag}
\affiliation{Centre for Condensed Matter Theory, Department of Physics, Indian Institute of Science, Bangalore 560012, India}

\author{Tejender S Thakur}
\affiliation{Biochemistry and Structural Biology Division, CSIR–Central Research Drug Institute, Lucknow 226301, India}

\author{Subramanian Yashonath}
\affiliation{Solid State and Structural Chemistry Unit, Indian Institute of Science, Bangalore 560012, India}

\author{Bhalamurugan Sivaraman}
\affiliation{Physical Research Laboratory, Navrangpura, Ahmedabad 380009, India}

\author{Prabal K. Maiti}
\email{maiti@iisc.ac.in}
\affiliation{Centre for Condensed Matter Theory, Department of Physics, Indian Institute of Science, Bangalore 560012, India}

\title[]
  {Deciphering Interstellar Ice Morphology: Atomistic Simulations Reveal the Complex Behavior of Ethanethiol}

\begin{document}







\begin{abstract}
Ethanethiol (C$_2$H$_5$SH), a molecule detected in the interstellar medium (ISM), indicates the rich chemistry involving sulfur atoms.
However, its behavior at low temperatures remains elusive, particularly the reported transition from an amorphous phase to a crystal. 
This study employs classical molecular dynamics (MD) simulations to reproduce the liquid-state properties of ethanethiol and to simulate the initial amorphous state of ethanethiol films deposited on a KBr substrate. 
The amorphous ethanethiol did not show spontaneous crystallization upon increasing temperature.
Also, ethanethiol ice crystals exhibit melting behavior on KBr substrate at elevated temperatures. 
Our MD simulations of thin ice samples do not show any signature reversible phase change. 
It will be interesting to continue this study with a thicker sample, which is beyond our current computational means.
These findings underscore the complexity of icy mantle morphology on cold ISM dust grains.
\end{abstract}

\section{Introduction}
\label{sec:1}
The formation of ethanethiol in the ISM likely occurs through a combination of two main processes. 
In frigid temperatures, the icy dust grain surfaces provide a platform for gas-phase molecules to stick and react. 
Here, atomic hydrogen (H) could interact with ethanethiol's precursor, C$_2$H$_5$S radical, to form the molecule \cite{muller2016exploring, cuppen2017grain, gorai2017search}. 
However, the efficiency of such reactions is bounded by the availability of free radicals that can react to form ethanethiol \cite{gorai2017search, hudson2024infrared}.
The major supply of these free radicals comes from another process that occurs in high-energy environments, such as shocks caused by supernovae or stellar winds where the gas-phase reactions become more prominent. 
Several such reaction mechanisms have been proposed so far by Gorai \emph{et al.} \cite{gorai2017search}.
A potential route in this scenario involves the interaction of methanethiol with free-floating carbon atoms (C) and radicals, leading to the formation of ethanethiol \cite{rodriguez2021thiols}.
Observations also suggest a significant depletion of sulfur on interstellar dust grains, potentially bound as $S_8$ or other allotropes \cite{shingledecker2020efficient}. 
Thus, shock-induced sputtering of these grains has the potential to liberate a substantial portion of sequestered sulfur, making it available for gas-phase reactions that could lead to the formation of ethanethiol and other sulfur-bearing species \cite{rodriguez2021thiols}. 
Both processes highlight the diverse chemical nature within the interstellar medium.

Studies suggest that ethanethiol can undergo reversible phase transitions between amorphous and crystalline phases temperature \cite{pavithraa2017qualitative}. 
This highlights the dynamic nature of interstellar ices and their potential role in complex molecule formation.
The result of this laboratory-based experiment was more astonishing as it also showed the dependence on the thickness and deposition temperature on the amorphous to crystalline phase transition.
The adsorbed ethanethiol layer on a potassium-bromide (KBr) substrate remains amorphous as the temperature was increases from $4$ K to $100$ K. 
However, upon further heating to a temperature range between 110 K and 120 K, the adsorbed phase undergoes a transition to a crystalline state.
Beyond 125 K and up to 130 K, the study suggests a potential return to an amorphous state for the adsorbed phase.
This entire observation was based on the IR spectra of the ethanethiol molecules.

Prior computational investigations of ethanethiol have primarily concentrated on the liquid phase and identification of its various vibrational modes\cite{miller2010overtone, phillips2022infrared, dodda20171}. 
The potential formation of ethanethiol within the icy mantles of cold interstellar regions suggests a need for further evaluation of current simulation studies to ensure their applicability to such environments.
The recent detection of ethanethiol in the interstellar medium (ISM) highlights the paucity of readily available data on its crystal structure in the scientific literature.

Therefore, the main aim of this work is to have a microscopic understanding of the reversible phase changes as seen in the experiments. 
This first requires the prediction of the possible crystal structure, which is currently missing in the literature.
We have first predicted a possible crystal structure of ethanethiol ice and used it to understand the thermodynamics of pure ethanethiol in the temperature range of 10K-140K.
Then, we used this model to study the properties of deposited ethanethiol molecules on KBr substrate as done in the experiment of Pavithraa \emph{et al.} \cite{pavithraa2017qualitative}. 

\section{Method}
\label{sec:2}
In the absence of any reported crystal structure of ethanethiol, we performed a crystal structure prediction (CSP) for this molecule to use it as the initial configuration for our simulation.
At first, a rigid CSP is performed with COMPASS force field (FF) \cite{sun1998compass}. 
The top 10 predicted structures were then taken for subsequent dispersion corrected (B3LYP-D) periodic-DFT-based energy re-ranking \cite{GrimmeDFT-D}.
The final top-ranked structure was considered as the unit cell of the ethanethiol ice structure.

The CSP crystal structure has a triclinic (P-1) unit cell, as shown in Fig. \ref{image_config_density_bulk}(a). 
The unit cell lengths are $a = \SI{6.61}{\angstrom}$, $b = \SI{6.09}{\angstrom}$, $c = \SI{4.43}{\angstrom}$, and the associated angles are $\alpha = \SI{87.31}{\degree}$, $\beta = \SI{85.44}{\degree}$, $\gamma = \SI{74.18}{\degree}$.
To study the properties of bulk ethanethiol, an initial configuration of bulk ethanethiol ice was generated by replicating this unit cell in all three directions to construct a supercell ($5 \times 5 \times 5$) that consists of $250$ molecules. 
A supercell of ($n_{x} \times n_{y} \times n_{z}$) denotes the number of replicas of unit cells placed along the $x$, $y$, and $z$ directions respectively; thereby denoting different sizes of crystal.
We used OPLS-AA FF parameters to describe the bonded interactions and non-bonded Lennerd-Jones interactions \cite{jorgensen2005}.
The ESP charges from CSP calculation were reused for electrostatics.
The initial bulk crystal was energy-minimized using the steepest descent method.
This was followed by heating the system at a rate of $\SI{0.05}{\kelvin \per\pico\second}$, and then performing NPT equilibration at temperatures $\SI{10}{\kelvin}$-$\SI{298}{\kelvin}$ at atmospheric pressure.
To maintain constant temperature (T) and pressure (P) throughout the simulation, Berendsen thermostat and barostat were employed with temperature coupling constant of $1$ ps and pressure coupling constant of $0.1$ ps respectively \cite{bussi2007canonical, berendsen1984molecular}. 
We used an integration time-step of $\SI{1}{\femto\second}$ and used periodic boundary conditions (PBC) in all three directions.
Short-range Lennard-Jones (LJ) and electrostatic interactions were treated with a cutoff distance of $\SI{10}{\angstrom}$ with smooth switching from $\SI{8}{\angstrom}$. 
Long-range electrostatics were handled using the Particle Mesh Ewald (PME) method with fourth-order interpolation \cite{darden1993particle, essmann1995smooth}.
The neighbour list was updated every 10 time-steps using the velocity Verlet algorithm.
All the simulations were performed using GROMACS MD engine \cite{abraham2015gromacs}.

To study the reversible phase transition of ethanethiol as reported by Pavithraa \emph{et al.} \cite{pavithraa2017qualitative}, we conducted deposition simulation, and also adsorption simulation of ethanethiol ice over KBr substrate.
The `vacuum deposition' of ethanethiol over KBr was carried out using PYTHINFILM - a python package that works in tandem with GROMACS \cite{abraham2015gromacs, stroet2022pythinfilm}.
The deposition protocol is similar to a previous work from our group further details of the procedure can be found in reference \cite{ramachandran2024amorphous, stroet2022pythinfilm}.
The potassium (K) and bromine (Br) atoms of KBr were parameterized using FF parameters developed by Barbosa \emph{et al.} \cite{fuentes2018potassium, nag2023influence}.

Our study of adsorbed ethanethiol ice over KBr substrate was done by placing ethanethiol ice crystal of supercell size ($6$x$6$x$6$,) ($6$x$6$x$12$), ($6$x$6$x$18$), ($6$x$6$x$24$) on the substrate.
To place the ice at an appropriate distance, minimum energy criteria in orientation and distance were used.
The minimum energy orientation is the same as that of the initial unit cell found by CSP, and the energetically minimum distance was $\SI{2.546}{\angstrom}$ between the surface of the substrate and ice.

Density of states (DoS) was calculated using Two-phase thermodynamics (2PT) \cite{lin2003two, lin2010two}.
For the 2PT calculations, NVT run of $\SI{20}{\pico\second}$ was performed with high-frequency dumping of position, velocities, and energies of atoms at every $\SI{2}{fs}$.

\section{Results and Discussion}
\label{sec:3}
\subsection{Bulk Ethanethiol}\label{section_pure_bulk_ethanethiol}

\begin{figure}[ht]
\centering
    \includegraphics[width=0.8\textwidth]{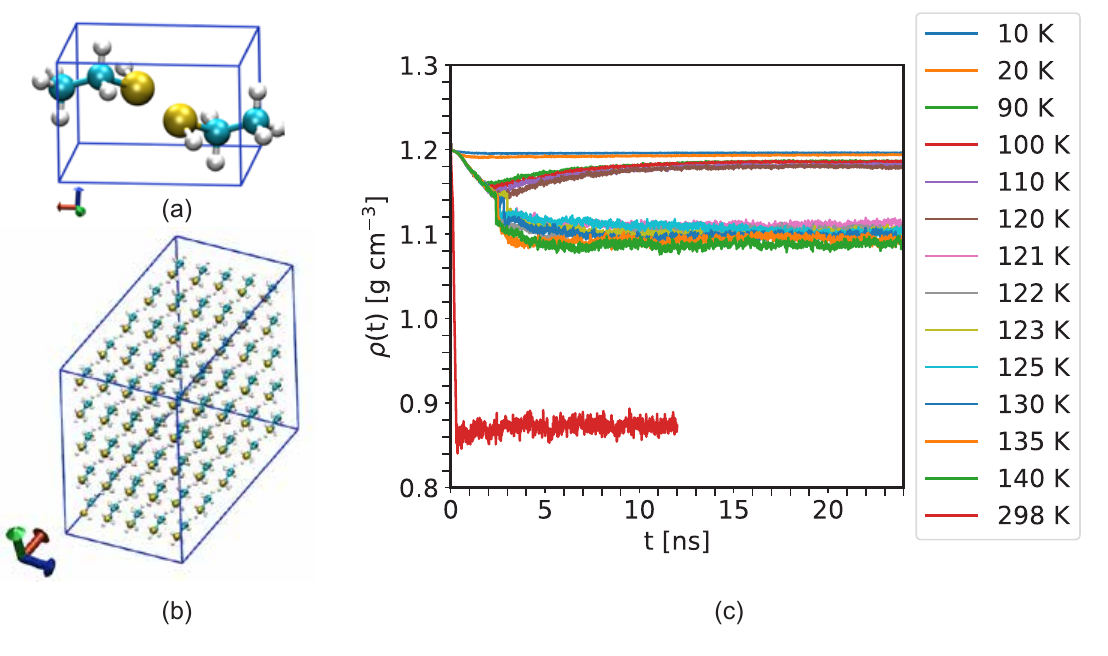}
  \caption{
  (a) Predicted crystal structure for ethanethiol. Sulfur, carbon, and hydrogen atoms are shown in yellow, cyan, and white colour, respectively. Red arrow, green arrow and blur arrow points to $x$, $y$ and $z$-axis respectively.
  (b) Crystal supercell ($5$x$5$x$5$) consisting of 250 ethanethiol molecules after $\SI{24}{\nano\second}$ simulation at $\SI{10}{\kelvin}$.
  (c) The time evolution of the density of bulk ethanethiol crystal, heated from $\SI{10}{\kelvin}$ to $\SI{140}{\kelvin}$ at a rate of $\SI{0.05}{\kelvin \per\pico\second}$ in ambient pressure. 
  The density equilibration shows that ethanethiol crystal undergoes a phase transition at $\SI{120}{\kelvin}$. 
  For reference, the density of ethanethiol upon direct heating to room temperature ($\SI{298}{\kelvin}$) is shown in red.
  }
  \label{image_config_density_bulk}
\end{figure}

\begin{figure}[ht]
\centering
\includegraphics[width=\textwidth]{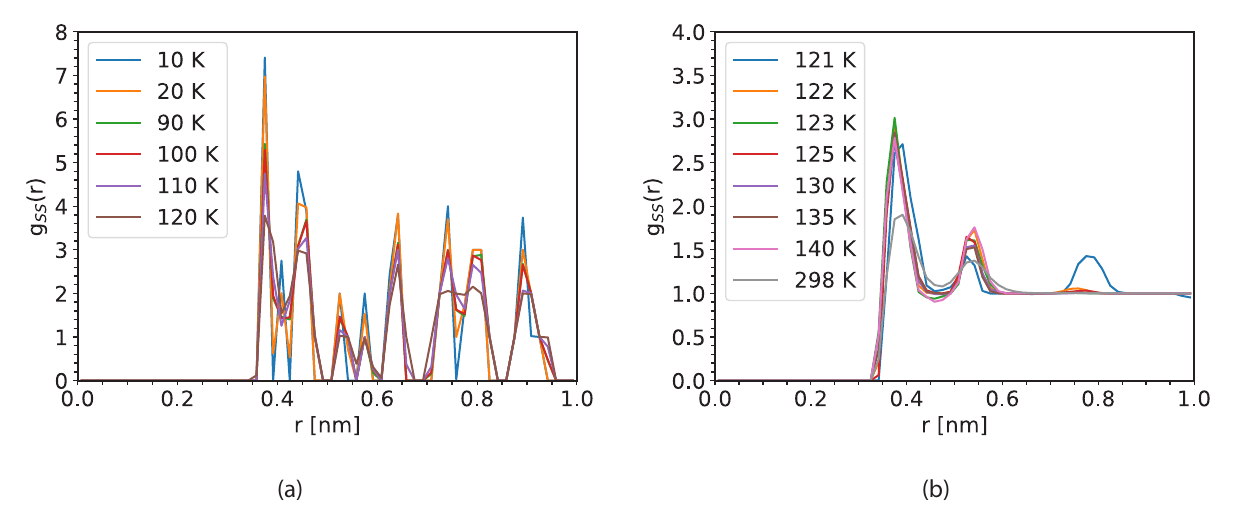}
\caption{
RDF of bulk ethanethiol at various temperatures as the crystal sample was heated to higher temperatures to the range where it shows the first change in phase.
The first occurrence of phase change happens at $\SI{121}{\kelvin}$ when the RDF reduces to only 3 peaks of smaller amplitudes.
}
\label{image_rdf_bulk}
\end{figure}

The low temperature stability of ethanethiol crystal structure was studied on a supercell of ($5 \times 5 \times 5$) dimension.
A snapshot of the initial configuration system is given in Fig. \ref{image_config_density_bulk}(b).
As the temperature increases, the density of the system changes.
The time evolution of this system density is plotted in Fig. \ref{image_config_density_bulk}(c).
It could be noted that as temperature crosses $\SI{120}{\kelvin}$, the density drops from $\SI{1.178}{\gram \per\centi\meter\cubed}$ to $\SI{1.139}{\gram \per\centi\meter\cubed}$, implying a phase change.
The density of the new phase is lower than the phase of the initial crystal, but it is much higher than the density of liquid ethanethiol at $\SI{298}{\kelvin}$ (shown in red in Fig. (\ref{image_config_density_bulk}(c))). 
Experimentally, the density of ethanethiol at $\SI{298}{\kelvin}$ is known to be $\SI{0.833}{\gram \per\centi\meter\cubed}$ \cite{haines1956purification, dodda20171}.
In our simulation, we get this density to be $\SI{0.872}{\gram \per\centi\meter\cubed}$. 
The phase change is also evident from the radial distribution function (RDF) as plotted in Fig. \ref{image_rdf_bulk}.

Another important quantity that is often used as a signature of crystal ethanethiol is the InfraRed (IR) spectra of S-H stretching vibration \cite{pavithraa2017qualitative, wolff1985vibrational, hudson2018infrared}.
Literature suggests that, in the crystalline phase, S-H vibration mode gives rise to a sharp peak, which effectively captures the phase of the sample.
We calculated DoS that can reveal the modes of vibrations in the IR frequency domain of ethanethiol spectra. 

\begin{figure}[ht]
\centering
\includegraphics[width=\textwidth]{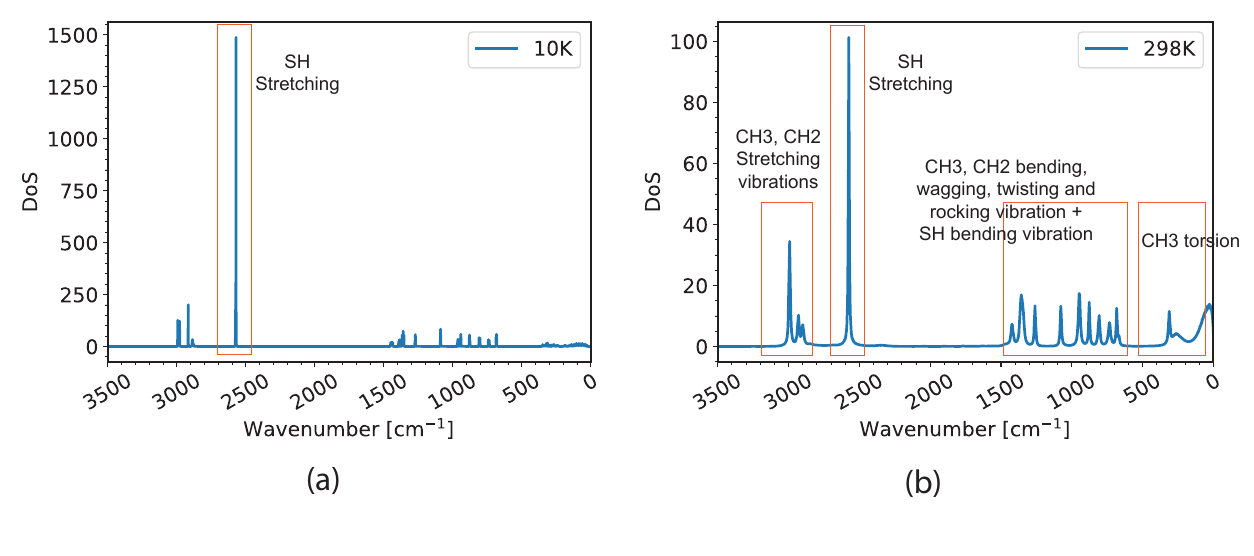}
    \caption{DoS of ethanethiol molecules in (a) crystal phase at 10K, and in (b) liquid phase at 298K.
    The vibration corresponding to the major peaks are marked where the tallest peak corresponding to $\SI{2570}{\per\centi\meter}$ indicates the S-H mode of stretching vibration.
  }
  \label{image_dos_comparison_10K_298K}
\end{figure}

Figure \ref{image_dos_comparison_10K_298K} shows the full DoS of ethanethiol at $\SI{10}{\kelvin}$ and $\SI{298}{\kelvin}$.
The major peaks in the DoS plot can be mapped to different modes of vibration and are marked in the liquid DoS plot of $\SI{298}{\kelvin}$.
The S-H peak can be identified as the highest non-overlapping peak at $\SI{2570}{\per\centi\meter}$.
The amplitude of this peak can be visually distinguished between the solid and liquid phases.
In the crystal phase at $\SI{10}{\kelvin}$, the S-H peak amplitude is approximately an order of magnitude higher than its value in the liquid phase at $\SI{298}{\kelvin}$.

\begin{figure}[ht]
\centering
  \includegraphics[width=0.8\textwidth]{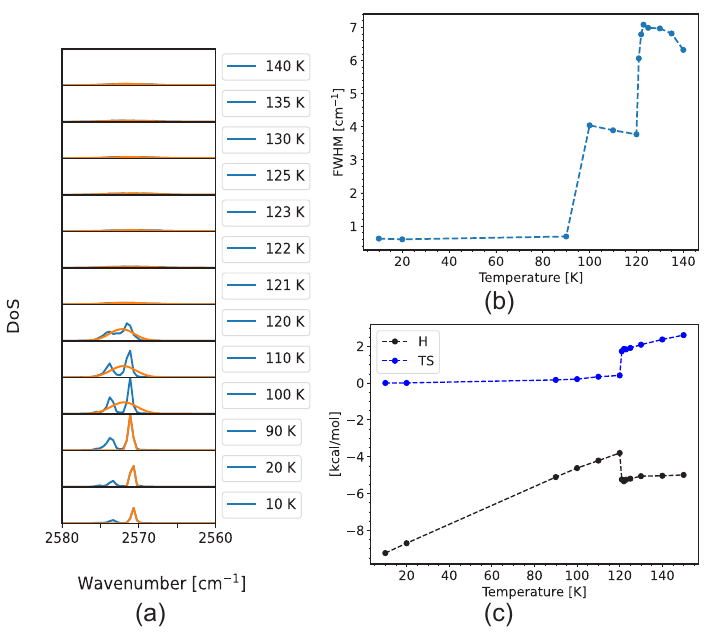}
  \caption{
  (a) The spectra of S-H stretching vibration mode in the DoS of bulk ethanethiol as system was heated from low temperature crystal configuration at $\SI{10}{\kelvin}$ to higher temperatures.
  The blue line shows the actual DoS plot and the orange line shows the Gaussian fitting.
  (b) The measured FWHM of the peak.
  (c) The enthalpy (black) and the temperature-weighted entropy (blue) with temperature. 
  All the plot captures the abrupt phase change of ethanethiol at $\SI{121}{\kelvin}$.
  }
  \label{image_bulk_IR_FWHM_TS}
\end{figure}

Figure \ref{image_bulk_IR_FWHM_TS} (a) shows this peak (enlarged view) as a function of temperature.
Fitting with a Gaussian function allows us to measure the peak details.
In the experiment, this S-H peak position is at $\SI{2528}{\per\centi\meter}$ \cite{pavithraa2017qualitative}.
In our simulation, we get this peak at $\SI{2570}{\per\centi\meter}$ therefore.
The full width at half maxima (FWHM) of this peak is plotted in Fig. \ref{image_bulk_IR_FWHM_TS} (b) which shows that at $\SI{121}{\kelvin}$ the FWHM shows a sudden increase by around $\SI{7}{\per\centi\meter}$.
This quantity is related to experimentally obtained IR spectra with some parameter rescaling.
Therefore, this confirms the phase change of ethanethiol from crystal to non-crystalline form at $\SI{121}{\kelvin}$ in our simulation.

In Fig. \ref{image_bulk_IR_FWHM_TS} (c), we have plotted the temperature-weighted entropy (TS) and enthalpy (H) of the system with temperature.
The discontinuity in both thermodynamic quantities at $\SI{121}{\kelvin}$ implies a phase change.
The rate of change of H and TS differs characteristically before and after the change of phase.
At $\SI{121}{\kelvin}$, $\Delta H$ is $\SI{1.44}{\kcal \per\mol}$ which can be compared to its experimental counterpart of $\SI{1.19}{\kcal \per\mol}$ \cite{mccullough1952ethanethiol}.

At $\SI{121}{\kelvin}$, we calculate the jump in enthalpy, which gives us the heat of fusion as $\SI{1.44}{\kcal \per\mol}$.
Again, at the liquid-gas phase boundary, the heat of vaporization can be computed using eq (\ref{equation_heat_of_vapourization}) as prescribed in Ref. \cite{wang2011application}:

\begin{equation} \label{equation_heat_of_vapourization}
\Delta H_{vap}(T) = E^{potential}_{gas}(T) - E^{potential}_{liquid}(T) + RT
\end{equation}

Using eq \ref{equation_heat_of_vapourization} we get the heat of vaporization from our simulation at $\SI{298}{\kelvin}$ to be $\SI{8.09}{\kcal \per\mol}$. 
Both the heat of fusion and heat of vaporization can be compared with experimentally known values and are summarized in Table \ref{table_comparison_bulk_properties}.

\begin{table}[ht]
\centering
\caption{Comparison of experimental properties of pure ethanethiol in literature vs. our simulations}
\begin{tabular}{|l|c|c|c|}
\hline
 Property &  Exp. &  Ref &  This work \\ \hline
 Density [\SI{}{\gram \per\centi\meter\cubed}] & 0.833 (298 K) & \cite{dodda20171} & 0.872 \\ \hline
 Heat of fusion [\SI{}{\kcal \per\mole}] & 1.19 (125.26 K) & \cite{mccullough1952ethanethiol} & 1.44 (121 K) \\ \hline
 Heat of vaporization [\SI{}{\kcal \per\mole}] & 6.35 (298.15 K) & \cite{dodda20171} & 8.09 (298 K)\\ \hline
 S-H stretching mode [\SI{}{\per\cm}]& 2528 & \cite{pavithraa2017qualitative} & 2570 \\ \hline
\end{tabular}
\label{table_comparison_bulk_properties}
\end{table}

This study on bulk ethanethiol and its first phase transition temperature helps us understand the temperature range over which it can sustain crystallinity in our simulation.
In the next sections, we use these results to understand the experimental observation of reversible phase transition in ethanethiol on substrate from our simulation perspective.

\subsection{Deposited Ethanethiol on KBr Substrate}
\label{section_deposited_ethanethiol}
In the experiment, ethanethiol deposited on the KBr substrate showed reversible phase change with temperature. 
However, this phase change was observed only for the `thick' samples $(\sim \SI{200}{\micro\meter})$ and not for the `thin' samples $(\sim \SI{200}{\nano\meter})$.
The thickness of the `thick' sample is prohibitively large; therefore, through MD simulation, we probe into the `thin' sample range.

\begin{figure}[ht]
\centering
\begin{subfigure}[b]{0.4\textwidth}
    \includegraphics[width=\textwidth]{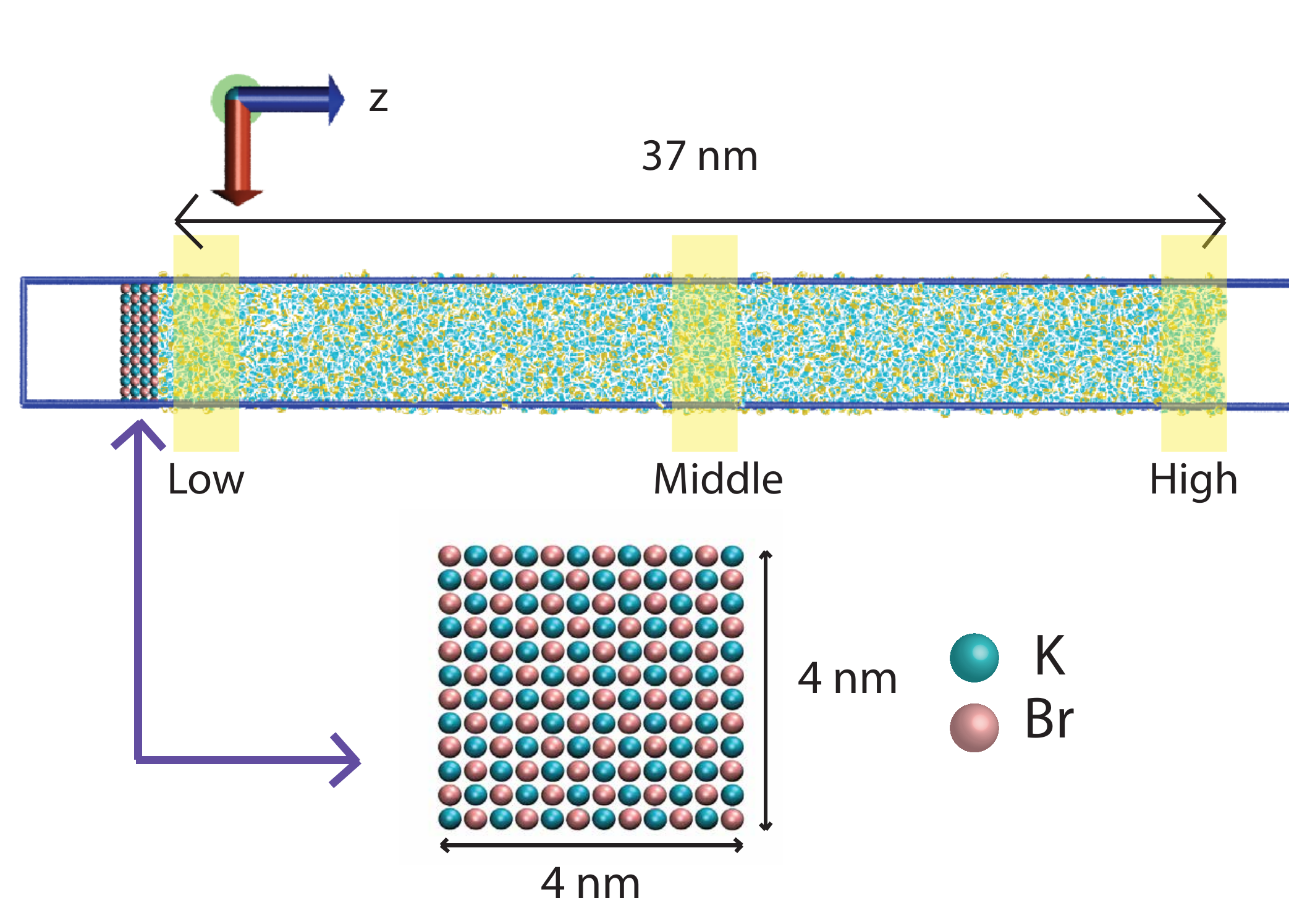}
     \caption{}
    \label{image_marked_regions_of_initial_deposition_10K}
\end{subfigure}
\begin{subfigure}[b]{0.4\textwidth}
    \includegraphics[width=\textwidth]{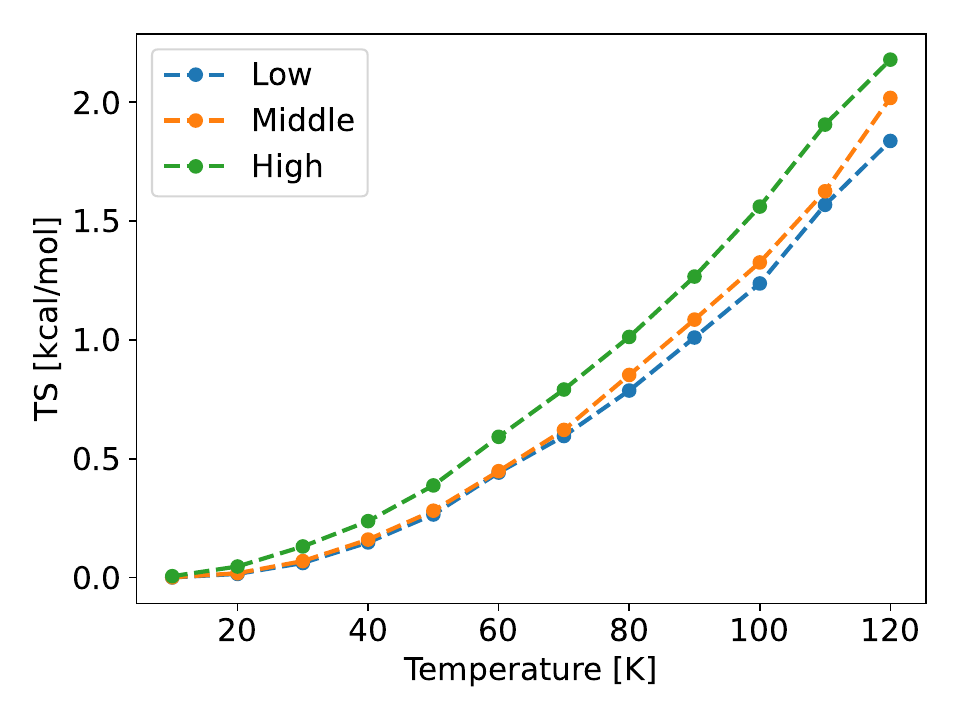}
    \caption{}
    \label{image_deposition_entropy_region}
\end{subfigure}
    \caption{
    (a)Deposited ethanethiol molecules over KBr substrate with marked regions that were analyzed independently.
    Deposited sample measure to $\SI{37}{\nano\meter}$, and each of these regions are of $\SI{3}{\nano\meter}$. 
    There is also a $\SI{20}{\nano\meter}$ vacuum between the deposited sample and the box ceiling along $z$.
    (b) Temperature-weighted entropy of the deposited molecules in the respective regions as a function of temperature.
    Compared to `Low' region molecules, the `High' region molecules show a greater rate in the increase of disorderliness with temperature.
    }
  \label{}
\end{figure}

Here, we deposit 5000 ethanethiol molecules on KBr substrate at $\SI{10}{\kelvin}$.
This substrate, of $4$x$4$ $\SI{}{\nano\meter \squared}$ area, have its atoms restrained to their initial coordinates with a force of force constant $\SI{1000}{\kilo\joule \per\mol}$ as shown in Fig. \ref{image_marked_regions_of_initial_deposition_10K}.
Molecules were released from a height of $\SI{6}{\nano\meter}$ and with an average velocity of $\SI{0.6}{\nano\meter \per \pico\second}$ towards the KBr surface.
A vacuum length of $\SI{20}{\nano\meter}$ was kept from the insertion distance to the box height to prevent periodic interaction along the $z$-direction.
The rate of heating, and equilibrium simulation settings were kept the same as that of pure bulk ethanethiol simulations described before.
At each temperature, simulations were carried out for $\SI{100}{\nano\second}$.

These deposited molecules on the KBr substrate were initially found to be in an amorphous phase, as shown in Fig. \ref{image_marked_regions_of_initial_deposition_10K}.
This system, thus created, was heated and simulated at higher temperatures in NVT ensemble (till $\SI{120}{\kelvin}$, the simulation temperature of crystal to non-crystalline transformation), which we then analyzed based on regions that are of equal size ($\SI{3}{nm}$): the `Low', `Middle', and `High'.
Figure \ref{image_marked_regions_of_initial_deposition_10K} marks these regions of the system.
The `Low' region stands for the lowermost part of the deposited sample, which is in close proximity to the substrate atoms.
The `Middle' region is defined with a sufficient buffer zone separating it from the top vacuum and the bottom substrate.
The `High' region is at the top of the deposited sample, which is at the boundary of the vacuum region.
In these individual regions, we analyze the system in tandem with experiments where crystallinity is detected through IR spectra.

\begin{figure}[ht]
\centering
  \includegraphics[width=\textwidth]{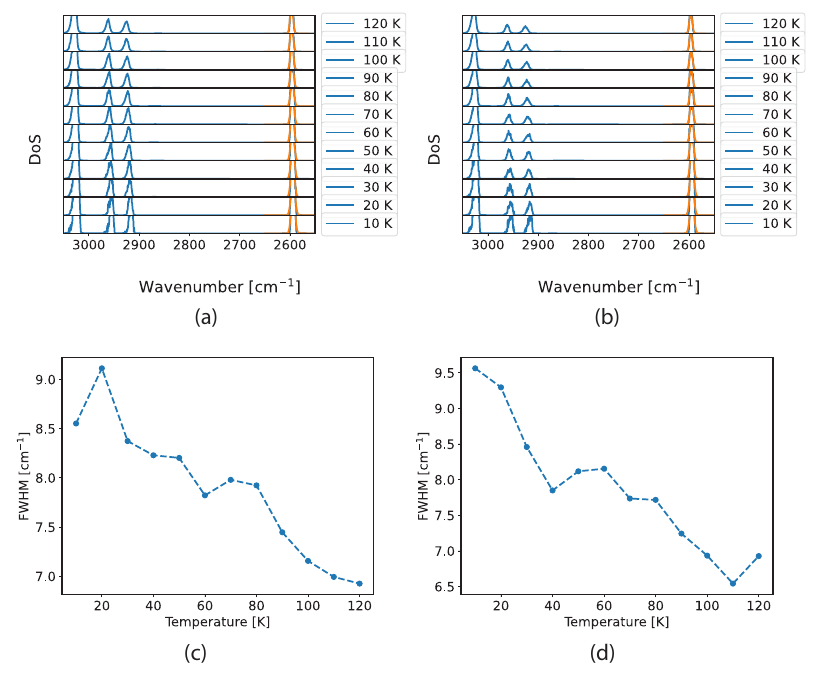}
  \caption{
  Spectra of ethanethiol in (a) `Middle' (b) `High' region of the deposited sample as a function of temperature.
  The FWHM of the S-H peak as a function of temperature is given in (c) `Middle' (d) `High'
  }
  \label{image_merged_deposition_IR_FWHM_region}
\end{figure}

We perform DoS calculations in these regions.
The representative spectra for the `Middle' and `High' are plotted in Fig. \ref{image_merged_deposition_IR_FWHM_region} (a), (b); the `Low' region which showed similar behaviour.
Fitting Gaussian on the S-H peak we monitor the FWHM of this peak in the individual regions that is plotted in Fig. \ref{image_merged_deposition_IR_FWHM_region} (c), (d).
Compared with bulk FWHMs (given in Section \ref{section_pure_bulk_ethanethiol}), it can be confirmed that the deposited sample (at $\SI{10}{\kelvin}$) is in amorphous phase.
With increasing temperature, there is a declining trend of FWHM in all the regions of the sample.
Therefore, deposited amorphous ethanethiol tends to achieve towards crystallinity with increasing temperature from $\SI{10}{\kelvin}$ to $\SI{120}{\kelvin}$.
However, the total reduction of FWHM from $\SI{10}{\kelvin}$ to $\SI{120}{\kelvin}$ is less than $\sim \SI{20}{\percent}$ which is far from a crystalline sample.

As a representative thermodynamic property, we calculated the TS of the molecules in these regions, and it is plotted in Fig. \ref{image_deposition_entropy_region}.
The plots show that for all the regions in the sample, entropy rises with temperature, and there is no abrupt jump that could be associated with phase change.
We also see that the rate of increase in disorderliness with rising temperature is higher for the molecules in the `High' region than the `Low' region.
Thus effectively, the thin sample does not crystallize under increasing temperature even below its first phase transition point. 
The results could also mean that an extremely slow crystallisation process is taking place in our sample, a common bottleneck in most equilibrium simulations.
Therefore, in the next section, we probe deeper with adsorbed ethanethiol ice over the KBr substrate.

\subsection{Adsorbed Ethanethiol Ice on KBr Substrate}

To understand the experimental observation in the case of `thin' ice, here we adopted a different perspective. 
To facilitate this study, adsorbed ethanethiol crystals of varying thicknesses were placed on a KBr substrate to test its crystalline stability with time.

\begin{figure}[ht]
\centering
\includegraphics[width=0.8\textwidth]{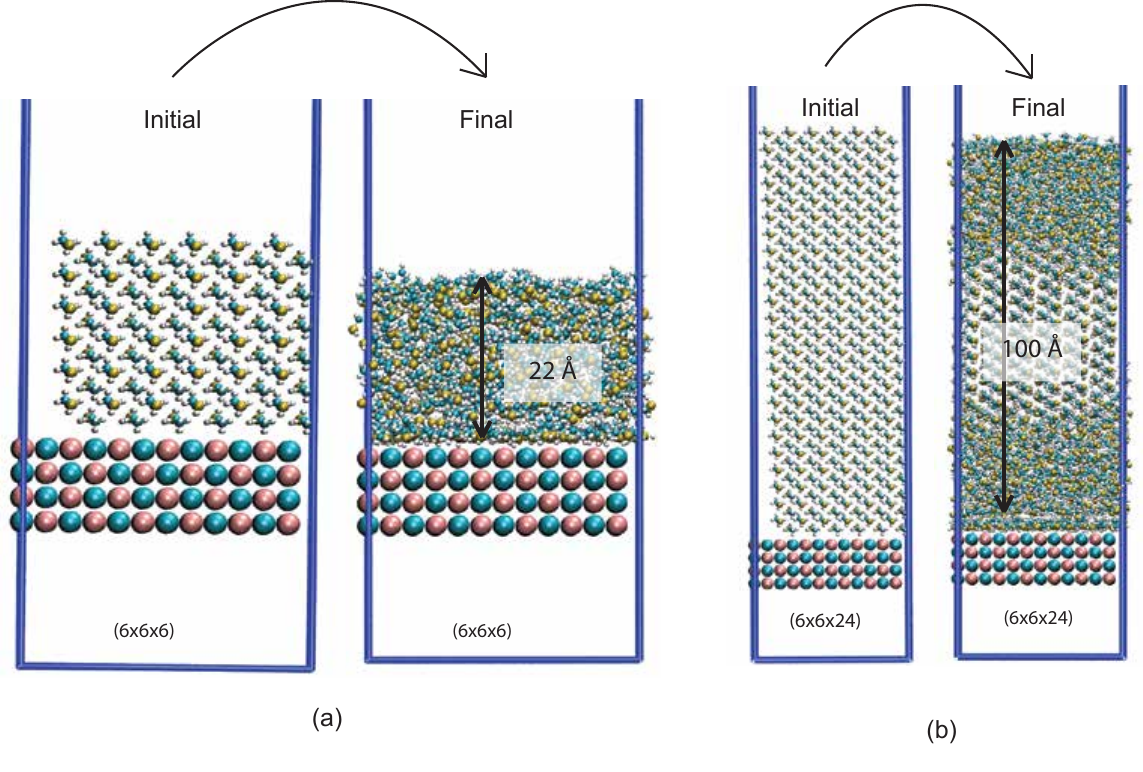}
\caption{
Initial and final configuration of adsorbed ethanethiol on KBr substrate for sample size of $\SI{22}{\angstrom}$ and $\SI{100}{\angstrom}$ sample height.
The final configuration is represents the system after $\SI{200}{\nano\second}$ simulation run.
}
\centering
\label{image_initial_final_config_adsorbtion_ethanethiol}
\end{figure}

It consists of ice crystals of different supercells: (6x6x6), (6x6x12). (6x6x18), and (6x6x24) as described in the method section.
Thus the initial height of adsorbed ethanethiol crystal to begin with varied from a thickness of $\SI{26.58}{\angstrom}$ for the smallest (6x6x6) system to  $\SI{106.32}{\angstrom}$ for the largest (6x6x24) system.
Here, since our objective is to check the crystal stability close to our bulk phase-change temperature $(\SI{121}{\kelvin})$, we slowly heated $(\SI{0.05}{\kelvin \per\pico\second})$ all the systems to $\SI{110}{\kelvin}$ and monitored their structure with time.

Our simulations revealed that the smallest system structure (6x6x6) exhibited melting within $\SI{5}{\nano\second}$. 
The (6x6x12) system underwent melting after a simulation time of $\SI{150}{\nano\second}$. 
Interestingly, even after $\SI{200}{\nano\second}$ of simulation, the 6x6x18 and (6x6x24) systems showed remarkable resilience against complete melting.
The initial and final configurations of the smallest (6x6x6) and largest (6x6x24) systems are given in Fig. \ref{image_initial_final_config_adsorbtion_ethanethiol}.
The larger system exhibited a gradual reduction in their crystal regions, starting from the exposed surfaces of both the vacuum and substrate.
These observations are evident from the snapshots in Fig. \ref{image_initial_final_config_adsorbtion_ethanethiol}.
The nature of melting for the remaining intermediate sample heights is similar.

\section{Conclusion}
\label{sec:4}

In this study, we addressed the knowledge gap in low temperature bulk ethanethiol properties using MD simulation.
Starting from bulk ethanethiol crystal structure prediction, we successfully reproduce the liquid-phase density of ethanethiol and deliver the heat-of-fusion and vaporization properties comparable to those of the experimental data.
Furthermore, we used the S-H stretching vibration peak from the DoS spectra as a fingerprint to identify crystalline ethanethiol ice.
It exhibited a significantly higher intensity in ordered crystals compared to room-temperature liquid molecules.

To understand the observed reversible phase change in thin ethanethiol films on a KBr substrate, we modelled deposited ethanethiol molecules. 
Consistent with experiments, the initial state resembled an amorphous phase, and no spontaneous crystallization occurred upon temperature increase.
To address the possibility of a slow crystallization process, we investigated the stability of pre-formed ethanethiol ice films on KBr at elevated temperatures (where the amorphous-to-crystal transition was observed experimentally). 
For all simulated `thin' film thicknesses (comparable to those used in the lab), the initial crystal structure degraded over time.
The breakdown of crystallinity initiated at both the KBr interface and the vacuum-facing surface, with the propagation rate likely dependent on the distance from the edges. 

This behavior highlights the possible role of hydrogen bonds in crystal formation, which is absent in ethanethiol due to the lack of highly electronegative atoms.
In conclusion, our simulations support the absence of the reported amorphous-to-crystal transition in `thin' ethanethiol films, aligning with the experimental observations.

\begin{acknowledgement}

We acknowledge funding through IISc STC project, Grant No. ISTC/PHY/PKM/500. J.M. thanks MHRD, India for financial support.

\end{acknowledgement}





\bibliography{refs}

\end{document}